\documentclass[]{article}

\usepackage{graphicx}
\usepackage{dcolumn}
\usepackage{bm}
\usepackage{bbm}
\usepackage[utf8]{inputenc}
\usepackage{amsfonts}
\usepackage{amsthm}
\usepackage{amssymb}
\usepackage{tikz}
\usepackage{physics}
\usepackage{comment}
\usepackage{xcolor}
\usepackage[caption=false,justification=justified]{subfig}
\usepackage{blindtext}
\usepackage{pgfplots,pgfplotstable}
\usepgfplotslibrary{polar}
\usepgfplotslibrary{colormaps}
\pgfplotsset{compat=newest}
\usepackage{hyperref}
\usepackage{xr-hyper}
\usepackage{footmisc}
\usepackage{cancel}
\usepackage{cleveref}
\usepackage[margin=1in]{geometry}
\usepackage{authblk}
\crefname{equation}{Eq.}{Eqs.}

\newcommand{\ii}{\mathrm{i}}

\author[1,2]{Fabrizio Sgobba}
\author[3]{Danilo Triggiani}
\author[3,4]{Vincenzo Tamma}
\author[5]{Paolo De Natale}
\author[2]{Gianluca Gagliardi}
\author[2]{Saverio Avino}
\author[1,2]{Luigi Santamaria Amato\thanks{luigi.santamaria@asi.it}}
\affil[1]{Italian Space Agency (ASI), Centro Spaziale ‘Giuseppe Colombo’, Località Terlecchia, 75100 Matera, Italy}
\affil[2]{Consiglio Nazionale delle Ricerche, Istituto Nazionale di Ottica (INO), Via Campi Flegrei 34, Comprensorio A.Olivetti, I-80078 Pozzuoli, Italy}
\affil[3]{School of Mathematics and Physics, University of Portsmouth, Portsmouth PO1 3QL, UK}
\affil[4]{Institute of Cosmology and Gravitation, University of Portsmouth, Portsmouth PO1 3FX, UK}
\affil[5]{Consiglio Nazionale delle Ricerche, Istituto Nazionale di Ottica (INO), Largo E. Fermi 6, I-50125 Firenze, Italy}

\title{Zeptosecond-scale single-photon gyroscope}
\date{}

\begin{document}

\maketitle

\begin{abstract} 
 This paper presents an all-fiber telecom-range optical gyroscope employing a spontaneous parametric down conversion crystal to produce ultra-low intensity thermal light by tracing-out one of the heralded photons. The prototype exhibits a detection limit on photon delay measurements of $249$ zs over a $72$ s averaging time and 26 zs in differential delay measurements at $t=10^4$ s averaging. 
 The detection scheme proves to be the most resource-efficient possible, saturating  $>99.5\%$ of the Cramér-Rao bound.
 These results are groundbreaking in the context of low-photon regime quantum metrology, paving the way to novel experimental configurations to bridge quantum optics with special or general relativity. 

\end{abstract}
\section{Introduction}
Fringe interferometry represents the most sensitive tool for optical metrology. The ability to discern tiny variations of interference patterns allows accurate measurements of environmental or sample parameters with a sensitivity scaling as the inverse frequency of the radiation source. Many different geometries, such as Michelson, Fabry-Pérot, Mach-Zehnder or Sagnac schemes to name a few, have been successfully investigated over time and employed in a wide variety of applications including high precision geometry \cite{yang2018review}, relativistic testing and gravitational wave detection (as in the LIGO/VIRGO experiments \cite{Abbott2016}), spectroscopy \cite{Hoghooghi2019} as well as accurate real time ranging \cite{Coddington2009}.

\noindent First demonstrated by Sagnac in 1913 \cite{sagnac1913ether}, the interferometric technique carrying his name is arguably one of the most robust among all interferometric configurations. 
Here, a single beam of light is separated into two counter-propagating beams and introduced in a closed loop, that may be arranged in different geometries  where the square or circular shape are most  common ones \cite{Arianfard2023}. An interference pattern arises whenever the  loop undergoes rotation, due to the difference in path-lengths  between the two counter-propagating beams.
Since both beams travel along the same path, such configuration is intrinsically insensitive to variations in temperature or environmental noise, such as mechanical vibrations, occurring on timescales  much larger than the round-trip time (as is usually the case for most environmental noise), thus making it particularly appealing for its intrinsic  robustness. 

Since the dawn of fiber-optic communication, Sagnac interferometry proved to be naturally suited to be implemented
 in all fiber-based setups, giving rise to the now widlely diffused fibre-optic gyroscope (FOG) interferometry \cite{Lefev2014}, a competitive alternative to the more complex  free-space ring laser gyroscope (RLG) \cite{Chow1985}, where an active Sagnac interferometer generates two measurable optical beating notes. These techniques became progressively more sensitive and reliable following the technological advances of the recent decades in optical fiber coupling as well as fiber-coupled optical sources and detectors, finding their application in so many different fields \cite{culshaw2005fiber}, from geodesy to hydrophony and geophony for military and civil applications, to precise ground positioning, that a comprehensive list would become a review on its own (see e.g. \cite{culshaw2005optical,lefevre2012fiber}).
In addition to FOG performances, which are rapidly improving, and to its fields of application, which are ever-expanding \cite{Arianfard2023}; a great deal of effort is being expended to produce increaslingly small FOG \cite{DellOlio2023}.
On the other hand, latest advances in single photon detection techniques \cite{Hadfield2023,dello2022advances} led to a strongly renewed scientific interest towards quantum optics and quantum interferometry beyond more direct application areas such as quantum communication, teleportation or quantum key distribution. For example, Hong-Ou-Mandel interferometry \cite{branning2000simultaneous,dauler1999tests} sensing capabilities when complemented by Fisher information analysis \cite{Lyons2018, triggiani2023freq} as well as delayed choice quantum erasers \cite{sgobba2023attosecond} led to attosecond-level resolution in delay measurements in low optical intensity regimes, paving the way for noninvasive biosensing and non-perturbing optical metrology.
Increasing applications of integrated devices based on Sagnac effect in quantum optics are expected in the near future as development of quantum sources \cite{Park2019} and sensors \cite{Grace2020} just to name a few.
\noindent In a quantum experiment aiming to highlight the interplay and coexistence of general relativistic effects with quantum mechanics, a sharp distinction has to be made between classical behaviours and quantum mechanics-related behaviours, as well as between general relativistic and Newtonian descriptions. If such boundary results more blurred when massive particles are involved \cite{Zych2012}, a fully optical experiment involving gravitational effects represents the ideal candidate due to the masslessness of the photon. Hence, the latest results in quantum metrology reported in \cite{Lyons2018, sgobba2023attosecond} together with challenging experiments adopting similar setups both for test of fundamental physics~\cite{Restuccia2019,Bertocchi2006} and for more practical applications~\cite{Fink2019,Zhou2003}, have fueled a wave of theoretical investigations on quantum-relativistic interface experiments both in free space and in fiber ~\cite{Rideout2012,Hilweg2017,Brady2021}.

\noindent Brady et al. \cite{Brady2021}, e.g., were able to estimate a lower bound on photon delay detection limit ($\sigma$) needed for the detection of the Sagnac effect due to Earth's rotation, using an interferometer of total area $\mathcal{A}$, via the figure of merit $F$, defined as $F=\frac{\sigma}{\mathcal{A}}$. The bound retrieved was $F \sim 10^{-15} \frac{\text{s}}{\text{km}^2}$, which means that attosecond-level ($\sim 10^{-18} \text{ s}$) delay sensing would require a total interferometer area $\mathcal{A}>1000 \text{ m}^2$.  

In this paper is presented what is, to the best of our knowledge, the first ultra-low photon regime delay sensor based on a 2 km fiber FOG with quadrupole-winding designed to operate in the telecom region (1550 nm) to minimise attenuation. The combination of highly performing near-infrared single photon detectors, together with quadrupole-winded FOG interferometry  and serrodyne modulation, not employed in other single photon FOG setups \cite{Fink2019,Restuccia2019}, led to establish a new limit both on the measurement of  the key quantity  
$F=\frac{\sigma_\tau}{\mathcal{A}} \sim 2\times 10^{-15} \frac{\text{s}}{\text{km}^2}$ at $t=72 $ s integration time and on the delay detection limit itself, obtaining for the first time an uncertainty as low as 249 zs in 72 sec integration time and  26 zs in differential
delay measurements, which, so far,  had never been achieved in photon counting regime.

\section{Experimental Setup}

\begin{figure}[ht]
\centering
\fbox{\includegraphics[width=0.5\linewidth]{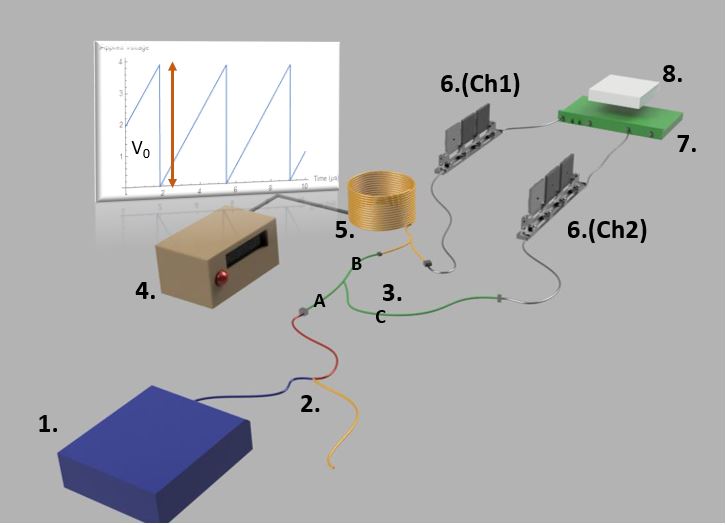}}
\caption{3D realistic sketch of the employed experimental setup. (1.) Twin photon source, (2.) polarising beam splitter, (3.) optical circulator, (4.) waveform generator, (5.) gyroscope, (6.) polarisation rotation paddles, (7.) Superconductive nanowires single photon detectors and (8.) Time tagging electronics}
\label{fig:stp}
\end{figure}
In the proposed setup the optical source is composed by a Twin Photons Source (TPS, 1. in Fig.\ref{fig:stp}), where a Continuous Wave (CW) diode laser, centred at $775$ nm, pumps a Periodically-Poled Lithium-Niobate (PPLN) crystal cut for type-II Spontaneous Parametric Down Conversion (SPDC), and specifically designed to maximise its  downconversion efficiency at the centre wavelength of  $1550$ nm, provided that its temperature is actively stabilised at $33.9 ^{\circ}$ C. 

\noindent Each correlated photon pair produced by the TPS is then separated via a polarising beam splitter (PBS, 2. in Fig.\ref{fig:stp}), so that it is always possible to select a single photon (having horizontal polarisation) from the pair. 

\noindent The selected photon enters in the fiber-coupled optical circulator from the input port $A$ (see 3. in Fig. \ref{fig:stp}).

\noindent The circulator allows light entering from the port A to come out of port B, whereas light entering from B comes out of port C. Port B is therefore connected to one of the inputs/outputs of a $2$ km-long FOG of total area $\mathcal{A}=125$ $\text{m}^2$. The FOG presents itself as a fiber-spool having average radius of $12.5$ cm (refractive index $n=1.471$) and a total volume of $\sim 6000$ $cm^3$. Therefore, the $2$ km fiber is wound up into $N \sim 2550$ coils. The interferometer is engineered in such a way that the forward and backward propagating paths undergo as similar environmental conditions as possible. In order to obtain this goal, the coils are wrapped in a quadrupole configuration, that ensures time-dependent temperature and strain variations affect in the same way both propagation directions. Embedded within the spool is present a fiber-coupled electro-optical modulator (EOM), tasked to provide a voltage-driven time-dependent phase shift between two propagating directions i.e., serrodyne modulation \cite{Serrodyne}. 

\noindent In the serrodyne modulation, the voltage has to be applied via a waveform generator (4. in Fig.\ref{fig:stp}) in the form of a saw-tooth wave having a period equal to twice the round-trip time inside the interferometer. In this specific case, the repetition rate results to be $\nu_{ramp}= 54.795$ kHz. In this experimental configuration, the total phase delay $\Delta \phi$ established between counter-propagating paths is directly proportional to the peak-peak voltage applied to the EOM ($V_0$ with reference to Fig. \ref{fig:stp}).

\noindent It is helpful to introduce the parameter $\tau=\frac{\Delta \phi}{\omega_0}$, which corresponds to a relative temporal delay between the counter-propagating paths that would generate the dephasing $\Delta\phi$ at central angular frequency $\omega_0$. $\tau$ is, in turn, proportional to the applied voltage $V_0$. Therefore, it can be expressed as
\[ \tau= \alpha V_0\]
Where we introduced the proportionality constant $\alpha$, the estimation of which has to be performed by calibration. For example, if a specific applied voltage $V_{0i}$  provides a total dephasing of $\Delta \phi=\frac{\pi}{2}$ for a $\lambda_0=1550$ nm photon between forward and backward propagating paths, 

\[
 \tau (V_{0i}) =\frac{\Delta \phi (V_{0i})}{\omega_0}=\frac{\pi}{2\omega_0}= \frac{\lambda_0}{4c}=1.294 \times 10^{-15} \text{ s} =1.294 \text{ fs}
\]
Where $c$ is the speed of light in vacuum.
Hence, 
\[
\alpha= \frac{\tau(V_{0i})}{V_{0i}}=\frac{1.294 \times 10^{-15}}{V_{0i}}\frac{\text{ s}}{\text{ V}}
\]
One of the two input/outputs of the gyroscope is then connected to the first channel (Ch1) of the detector, consisting in a 2-channels array of He-cooled superconducting nanowires single photon detectors (SNSPDs, 7. in Fig. \ref{fig:stp}). The second channel (Ch2) is connected to the port C of the optical circulator, so that light coming from the gyroscope and entering into port B can be collected in Ch2. Since SNSPDs efficiency is strongly dependent on impinging light polarisation, each channel is provided with a polarisation rotation paddle (6. in Fig. \ref{fig:stp}) in order to rotate the photon polarization to maximise the detectors efficiency.

\noindent Both channels are connected to a time tagging device (8. in Fig.\ref{fig:stp}) for data collection.

\section{Theoretical framework}
The pure photon-pair state generated via type-2 SPDC can be written as
\begin{equation}
\ket{\Psi}= \int_\mathbb{R} \dd\omega\ g(\omega) a_\mathrm{H}^\dagger(\omega)a_\mathrm{V}^\dagger(\omega_p-\omega) \ket{0}, 
\label{eq:SPDCState}
\end{equation}
where $\omega_p$ is the pump frequency, $g(\omega)$ is the joint spectral amplitude, and $a_\mathrm{H/V}^\dagger(\omega)$ denotes the photonic creation operator associated with a photon with frequency $\omega$ and polarisation H or V.
The effect of the polarizing beam splitter, selecting only the photons with vertical polarisation, is to trace out from \eqref{eq:SPDCState} the horizontally polarised photon.
The state injected in the FOG is thus mixed and reads
\begin{equation}
\rho=\int_\mathbb{R}\dd\omega\dd\omega'\ \rho(\omega,\omega')a^\dag_\mathrm{H}(\omega)\ketbra{0}a_\mathrm{H}(\omega')
\label{eq:InitialState}
\end{equation}
where 
\begin{multline*}
\rho(\omega,\omega')= \int_\mathbb{R}\dd\omega_1 \bra{0}a_\mathrm{H}(\omega) a_\mathrm{V}(\omega_1)\ketbra{\Psi}a^\dag_\mathrm{H}(\omega') a^\dag_\mathrm{V}(\omega_1)\ket{0}\\
= \abs{g(\omega)}^2\delta(\omega-\omega')
\end{multline*}
 is the matrix representation of $\rho$.

\begin{figure}[ht]
\centering
\fbox{\includegraphics[width=0.3\linewidth]{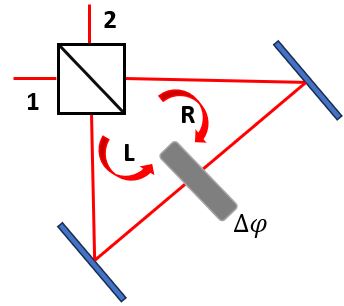}}
\caption{Schematic depiction of a Sagnac interferometer.}
\label{fig:Sagnac_Interf}
\end{figure}

For all extents, the FOG introduced in the previous section can be treated as a Sagnac interferometer, or a Mach-Zehnder interferometer where input and output beam splitters (BS) coincide, as shown in Fig. \ref{fig:Sagnac_Interf}

In the proposed setup, the photon enters from port $2$, with reference to Fig. \ref{fig:Sagnac_Interf}. 
Since, for all practical purposes, the photon polarisation state is irrelevant for the remainder of these calculations, for simplicity we will omit the subscript H from \eqref{eq:InitialState} and we replace it with a subscript $i=1,2,R,L$ indicating, again with reference to Fig. \ref{fig:Sagnac_Interf}, the channel in which the photon propagates.
The evolution of the operator $a_2^\dag(\omega)$, impinging on the beam splitter and undergoing the phase differential $\Delta\varphi$ first, and then exiting the interferometer, is given by
\[a^\dag_2(\omega)\to\frac{1}{\sqrt{2}}\left( a_R^\dagger(\omega)e^{\ii\Delta\phi} +
a_L^\dagger(\omega)\right)\to\]
\[\frac{1}{2}\left[a_2^\dagger(\omega)\left(e^{\ii\Delta\phi}+1\right) +
a_1^\dagger(\omega)\left(e^{\ii\Delta\phi} -1\right) \right].\]
The state at the output of the interferometer can thus be written as
\[
\rho_{\mathrm{out}}\!=\!\frac{1}{4}\int_\mathbb{R}\dd\omega\abs{g(\omega)}^2 \left[a_2^\dagger(\omega)\left(e^{\ii\Delta\phi}+1\right) +
a_1^\dagger(\omega)\left(e^{\ii\Delta\phi} -1\right)\right]
\]
\[
\ketbra{0} \left[a_2(\omega)\left(e^{-\ii\Delta\phi}+1\right) +
a_1(\omega)\left(e^{-\ii\Delta\phi} -1\right)\right],\]
where $\Delta \phi=\omega\tau$ is the relative dephasing between the two counter-propagating paths.

\noindent The probabilities to register a click at the first or second channel, respectively $P_1$ and $P_2$, can therefore be written as
\[
P_{2/1}=\int_\mathbb{R}\dd\omega\ \bra{0}a_{2/1}(\omega)\rho_{\mathrm{out}}a_{2/1}^\dag(\omega)\ket{0}\]
\[=\frac{1}{2} \left(1 \pm \int_\mathbb{R}\dd\omega\ |g(\omega)|^2 \cos(\omega\tau)\right).
\]
 For a photon having a Gaussian spectral distribution with a linewidth $\sigma_\omega$ and a center angular frequency $\omega_0$, described by a normalised spectral distribution 
\[ \abs{g(\omega)}^2=\frac{1}{\sqrt{2 \pi \sigma_\omega^2}}e^{- \frac{(\omega-\omega_0)^2}{2\sigma_\omega^2}},\]
the probability distribution reads
\begin{equation}
    P_{2/1}=\frac{1}{2} \left(1 \pm e^{-\frac{\sigma_\omega^2 \tau^2}{2}} \cos(\omega_0\tau)\right).
\label{probs}
\end{equation}
From \eqref{probs}, it follows that by measuring the counts events on each channel (which are proportional to the outcome probability $P_{1/2}$ once the dark counts are removed), it is possible to retrieve the value of the delay $\tau$ between propagating and counter-propagating probability amplitudes, and ultimately to retrieve the rotational angular velocity $\Omega$ of the reference frame to which the Sagnac interferometer is bound. In order to obtain the most sensitive possible experimental condition for the measurement of the parameter $\tau$ it results mandatory to evaluate the Fisher information, namely the variance of the score (defined as the gradient of the logarithm of the likelihood function)~\cite{kay1993statistical, Lyons2018, Harnchaiwat2020}.
As extensively discussed in \cite{kay1993statistical}, given a parameter $\theta$ conditioning the measurement outcomes $m \in M$ with a probability distribution $P(m|\theta)$, which expresses the probability of outcome $m$ given $\theta$, the Fisher information is defined as
\[
\mathcal{F}(\theta)= \sum_{m\in M} \frac{1}{P(m|\theta)} \left( \frac{\partial}{\partial \theta}P(m|\theta)\right)^2.
\]
The metrological importance of the Fisher information resides in the fact that it yields the maximum precision achievable in the unbiased estimation of $\tau$ with a given measurement scheme through the Cramér-Rao bound~\cite{Cramer1999}
\[
\sigma_\tau(N) \geqslant \frac{1}{\sqrt{N \mathcal{F}(\tau)}},
\]
where $\sigma_\tau(N) $ is the uncertainty associated with the estimation, and $N$ is the number of repetitions of the experiment, namely the total number of detected photons.
Hence, larger values of the Fisher information yield better sensitivities. 

\noindent Once the Fisher information and the number of detected photons have been determined, the degree of saturation of the Cramér-Rao bound (i.e. the quantity $S(\tau)=(\sqrt{N\mathcal{F}(\tau)}\sigma_\tau(N))^{-1}$, $S\leq1$) is a powerful metric to determine how efficiently the experimental setup is capable to extract information by each one of the detected photons. A fully saturated ($S(\tau)=1$) Cramér-Rao bound therefore qualifies the experimental procedure as the most efficient to estimate the chosen parameter, all other experimental conditions being equal.

\noindent The partial derivative squared of the probability function $P_{2/1}$ with respect to the conditioning parameter $\tau$ follows from~\eqref{probs} as
\begin{align*}
(\partial P_{2/1})^2&= \frac{1}{4}\left(\mp \frac{\sigma_\omega^2}{2} \tau e^{-\frac{\sigma_\omega^2 \tau^2}{2}}\cos(\omega_0\tau) \mp \frac{\omega_0}{2}e^{-\frac{\sigma_\omega^2 \tau^2}{2}}\sin(\omega_0\tau)\right)^2\\
 &=\frac{1}{4}e^{-\sigma_\omega^2 \tau^2}\left(\sigma_\omega^2 \tau \cos(\omega_0\tau) + \omega_0\sin(\omega_0\tau)\right)^2.
\end{align*}
The Fisher information of this system, as a function of the parameter $\tau$, reads therefore
\begin{equation}
\mathcal{F}(\tau)= e^{-\sigma_\omega^2 \tau^2} \frac{\left(\sigma_\omega^2 \tau \cos(\omega_0\tau) +\omega_0 \sin(\omega_0\tau)\right)^2}{1-e^{-\sigma_\omega^2 \tau^2}\cos^2(\omega_0\tau)}.
\label{fisher}
\end{equation}

\begin{figure}[h]
    \centering
    \fbox{\includegraphics[width=0.5\linewidth]{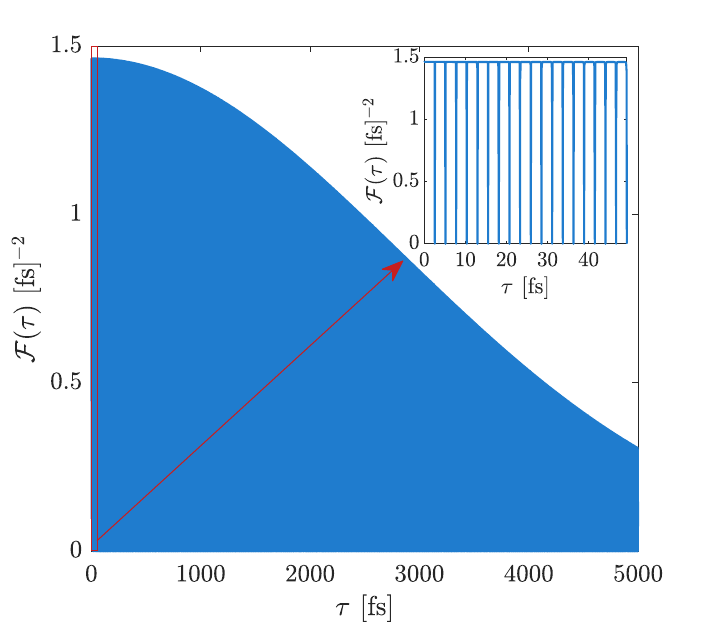}}
    \caption{Simulation of the Fisher Information $\mathcal{F}(\tau)$ for the system reported in Fig.\ref{fig:stp}, the inset shows $\mathcal{F}(\tau)$ in the first 50 fs.}
    \label{fig:FI}
\end{figure}

\noindent The linewidth of the SPDC source employed results to be $\sigma_\omega \approx 0.25$ THz, whereas $\omega_0= 2\pi\nu_0 \approx 1.21 \times 10^{3}$ THz. The Fisher information evaluated for a delay varying in the range $\tau \in [0, 5000\text{ fs}]$ is reported in \figurename~\ref{fig:FI}.

\noindent In low delays regime, where $\sigma_\omega\tau\ll 1$ and $\sigma_\omega^2\tau\ll \omega_0$, \eqref{fisher} becomes
\[ \mathcal{F}(\tau) \approx \omega_0^2\]
everywhere except for $\omega_0\tau= (2l+1)\pi$, $l \in \mathbb{N}$.









\section{Results and discussion}
\subsection{Gyroscope characterisation and sensor calibration}
\begin{figure}[ht]
\centering
\fbox{\includegraphics[width=0.5\linewidth]{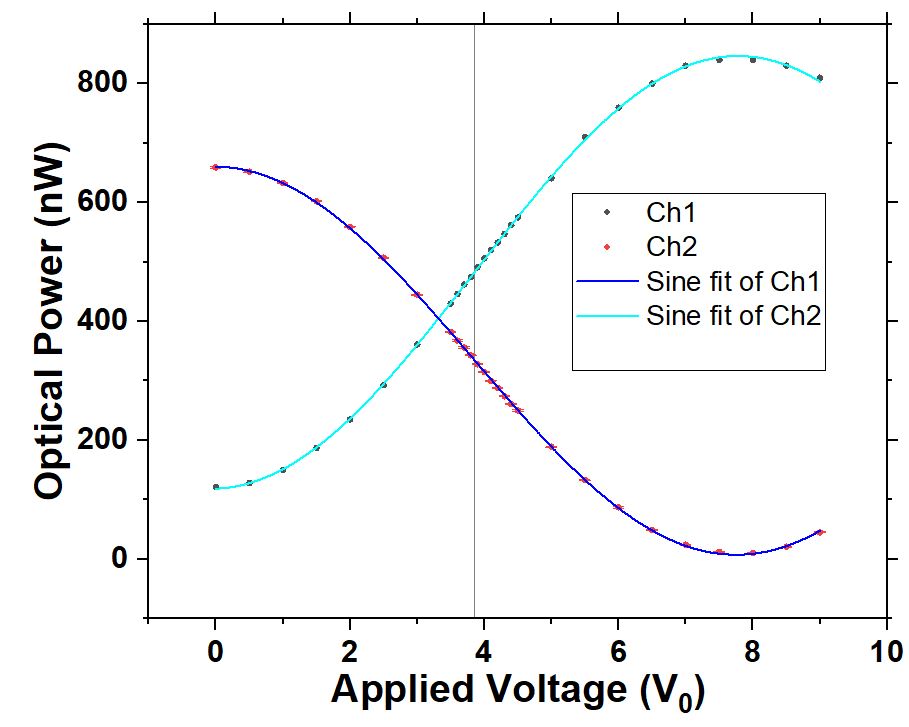}}
\caption{Optical power output of the FOG while employing a filtered incoherent bright source for Ch1 (in black) and Ch2 (in red), measured at a fixed applied Voltage (V). A reference line is reported for $V_0=3.8596$ V. The fits of Ch1 and Ch2 with the function $f_{fit}(V_0)$ are reported as blue and cyan lines respectively.}
\label{fig:SLED}
\end{figure}

In order to determine precisely the inflection point for the probabilities in \eqref{probs}, where $\Delta \phi_i=\frac{\pi}{2}$, we decided to operate in high intensity regime and we replace the twin photon source with a superluminescent light emitting diode (SLED) filtered at $1550$ nm with a bandwidth of $1$ nm. The output power has been then measured as a function of the voltage $V_0$ applied to the waveform generator by means of an optical power meter. Acquired data have been plotted and fitted with a sine function 
\[ 
f_{fit}(V_0)= f_0 + A\sin\left(\pi \frac{V_{0}- V_{0i}}{w}\right)
\]
for both gyroscope outputs (see Fig. \ref{fig:SLED}).
The values of the parameters $A,w,f_0,V_{0i}$ have been extracted from the plot in Fig. \ref{fig:SLED} and summarised in Table \ref{tab} with their respective fit errors. From these,  the final value for the inflection point has been retrieved via weighted average as  
\[
<V_{0i}>= 3.8596 \pm 0.0095 \text{ V}.
\]
Consequently, the proportionality constant $\alpha$ results
\[
\alpha = (3.35 \pm  0.03)\times10^{-16}\frac{\text{ s}}{\text{ V}}.
\]
\begin{table}[h]
    \centering
    \begin{tabular}{|c|c|c|c|c|}
        Channel &$V_{0i}$ (V) & $w$ (V) & $A$ (nW) &  $f_0$ (nW)\\
        \hline \hline
        Ch1 & $3.85 \pm  0.01$ & $7.84 \pm 0.04$ & $364 \pm 1$ & $482 \pm 1$ \\
        Ch2 & $3.93 \pm 0.03$ & $7.79 \pm 0.03$ & $327 \pm 1$  & $334 \pm 1$ \\
    \end{tabular}
    \caption{Parameters retrieved from fit}
    \label{tab}
\end{table}

\noindent It can be observed how, if the half-period $w$ results comparable within the fit error, the amplitude $A$ and the offset $f_0$ differ between the channels. The $\sim 10\%$ discrepancy in amplitude has to be traced back to unbalanced losses in the channels that cannot be attributed to the FOG or to the particular detector employed (since the output power on both channels for the bright-source setup has been measured with the same optical power meter), but arguably depends on the fibers and on the circulator. The residual offset on Ch1, $f_0-A \sim 120 $ nW,  can be attributed to back-reflections happening before the EOM and affecting the output connected to Ch1 itself as well as to non perfect visibility. Residual offset on Ch2 results instead comparable to the background noise seen by the power meter.

\noindent Fig.\ref{fig:SLED} shows, that for an applied voltage $V_0$ comprised between the values of $3.6$ V and $4.4$ V, both channels show linearity between applied voltage and optical power. Hence, this region results to be the best candidate to perform a calibration of the delay sensor.

The actual delay sensor calibration and the subsequent measurements have been therefore performed with the single photon setup shown in  Fig. \ref{fig:stp}.

\noindent For $V_0 \in [ 3.6 \text{ V}, 4.4 \text{ V}]$, each value of the applied voltage $V_0$ (and consequently delay $\tau$) will return a value of single count events on each channel ($C_{1/2}$) linearly dependent on $V_0$. 

\noindent The introduction of
\[
X_1= \frac{C_1}{C_1 + C_2} \qquad X_2=  \frac{C_2}{C_1 + C_2}
,\]
allows to renormalise the single count events by the sum of all tagged counts in order to rule out fluctuations coming from the optical pump.
The calibration curve can be written as ($\Delta X= X_1-X_2$)
\begin{equation}
    \Delta X= K_1 \tau  + K_2.
    \label{calb}
\end{equation}

\noindent To reduce the amount of dark count events, both the calibration and the measurements reported in the following sections have been performed after sunset. In these conditions, each channel experienced stable dark counts of $dc \sim 20-30 $ Hz, close to electronics-related dark counts ($\sim 10-20 $ Hz) and negligible with respect to the signal on each channel during the measurement ($C_1 \approx C_2 \sim 300-350 $ kHz). 

\noindent The calibration routine consisted in $100$ steps between the values of $V_{0a}=3.6$ V and $V_{0b}=4.4$ V. Each step has been acquired $10$ times with an acquisition time of $100$ ms in order to retrieve, for each step, an average value of $(X_1,X_2, \Delta X)$ and an error, evaluated propagating the error resulting from the standard deviation on $C_{1/2}$ of the $10$ acquisitions per step.

\begin{figure}[h]
\centering
\fbox{\includegraphics[width=0.5\linewidth]{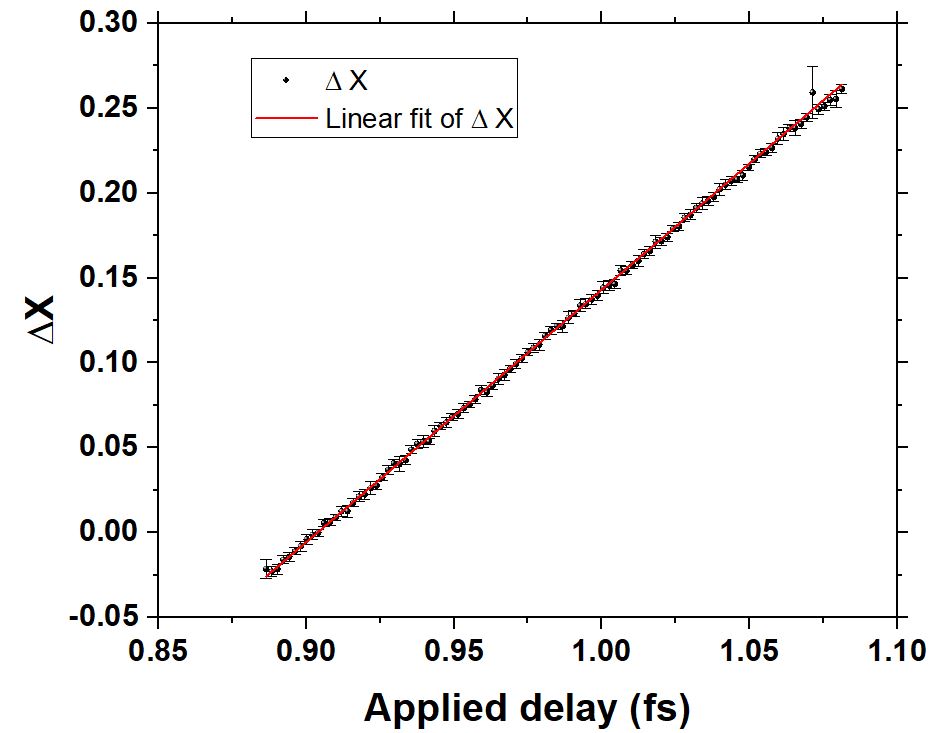}}
\caption{The experimental values of $\Delta X= X_1-X_2$, and the corresponding applied delays $\tau$  at which have been retrieved, fitted with a linear calibration curve}
\label{fig:calib}
\end{figure}

\noindent The calibration of the delay sensor is reported in Fig.\ref{fig:calib}, and the retrieved calibration parameters $K_1$, $K_2$ are reported in Table \ref{tab2}.

\begin{table}[h]
    \centering
    \begin{tabular}{|c|c|}
       $K_1$ (fs$^{-1}$) & $K_2$  \\
        \hline \hline
       $1.0937 \pm  0.0036$ & $-1.3432 \pm 0.0049$ \\
    \end{tabular}
    \caption{Parameters retrieved from calibration}
    \label{tab2}
\end{table}

\subsection{Long term measurement}

Right after the calibration we performed a $9$ h-long acquisition over night, using an integration time of $T=1$ s per point. The function generator commands were set to provide a sawtooth wave of peak to peak amplitude $V_0=3.86$ V.

\begin{figure}[h]
\centering
\fbox{\includegraphics[width=0.5\linewidth]{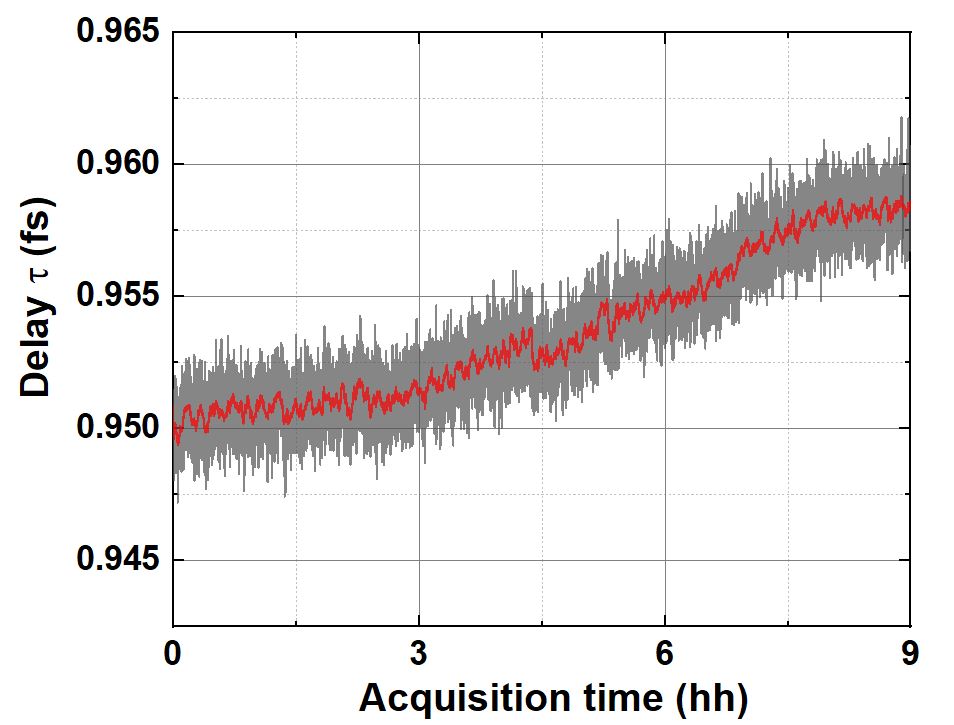}}
\caption{Long term acquisition with acquisition time expressed in hours (hh), $T=1$ s integration time, $V_0=3.86$ V applied voltage. In red, the same data smoothed by adjacent average at $72$ s integration time.}
\label{fig:longterm}
\end{figure}

\noindent A Python script was used to convert acquired single count events on each channel into delays $\tau$ by means of \eqref{calb}, employing the values of $K_1$, $K_2$ reported in Table \ref{tab2}. The whole acquisition is reported in Fig.\ref{fig:longterm}.

As can be observed from Fig. \ref{fig:longterm}, the system experiences a long term drift of $\sim 10$ as over $9$ h. Such drift can be associated to variations in temperature, affecting in particular the EOM piezoactuator embedded within the gyroscope.

\noindent In the medium-short term ($<10$ mins) the instabilities of the pump are still recognizable, even after the renormalisation. 

\noindent A time-domain frequency stability analysis allows both to assess the detection limit that the proposed delay sensor is capable to achieve at different integration times as well as to effectively investigate the different contributions to the noise affecting the signal. It is worth pointing out that, in this case, the detection limit (DL) of the sensor is intended as the measured time delay that can be achieved with a unitary signal to noise ratio (SNR), hence $DL(t)= \sigma(t)$.

\noindent For this reason, we performed an overlapping Allan (OA) deviation analysis of the long term measurement reported in Fig.\ref{fig:longterm}, where, given a set of $N$ datapoints $\{ x_1, ... x_N\}$ acquired at integration time $t_0$, the OA variance at a given time ($t=m t_0$) is defined as \cite{riley2008handbook}
\[
\sigma_{OA, x}^2 (t)= \frac{1}{2m^2(N-2m+1)} \sum_{j=1}^{N-2m+1} \left( \sum_{i=j}^{j+m-1} x_{i+m}-x_{i}\right)^2
.\]

\begin{figure}[ht]
\centering
\fbox{\includegraphics[width=0.5\linewidth]{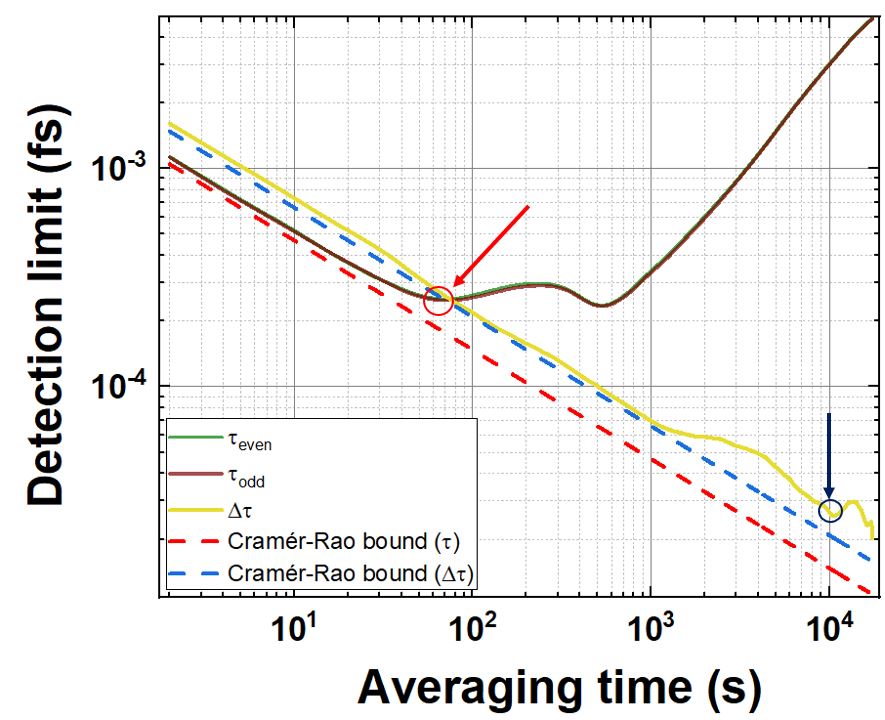}}
\caption{OAdev analysis of the long term measurement reported in Fig.\ref{fig:longterm}, sliced into even datapoints (in olive green), odd datapoints (in wine) and the deviation of their difference (in yellow). The red and blues dashed lines represent the Cramér-Rao bounds for the detection of $\tau$ and $\Delta \tau$, respectively, as a function of the averaging time, proportional to the detected photons. Red and blue arrows and circles highlight the detection limit for $\tau$ at $72$ s and $\Delta \tau$ at $10^4$ s.}
\label{fig:allan}
\end{figure}

Several quantum metrology applications \cite{Lyons2018,Harnchaiwat2020,sgobba2023attosecond} perform differential measurements on their conditioning parameter (in this case, $\tau$) to achieve better long term stability performances. To allow a more straightforward comparison with such literature, we decided to slice the data acquired and perform the OAdev analysis to even and odd data points separately (retrieving detection limits for $\tau_{even}$ and $\tau_{odd}$, respectively), de facto obtaining a $\Delta \tau=0$ differential measurement that represents the ultimate detection level achievable, once all noise sources (both coming from the source and the environment) are ruled out. The analysis has been plotted and reported in Fig.\ref{fig:allan}. 

\noindent 
\noindent The average count rate of both channels combined  during the long term measurement ($R$) reported in Fig.\ref{fig:longterm} corresponded to $R=631.6$ kHz. Therefore, since the $\tau_{even/odd}$ values are updated once every two seconds, the theoretical Cramér-Rao bound $CRB(\tau)$ derived from \eqref{fisher} becomes (red dashed line in Fig. \ref{fig:allan})
\[
\sqrt{\frac{1}{ \mathcal{F} N}} \approx  \sqrt{\frac{2}{ \omega_0^2  R t} } .
\]
\noindent For $t<60$ s, i.e. before medium term oscillations and long term drifts, the proposed system proves to saturate up to $93-95 \%$ of the Cramér-Rao bound. In particular, from the OA deviation analysis in Fig.\ref{fig:allan} emerges that the  detection limit on $\tau$ (red circle) is achieved for $t_{DL}=72$ s, in correspondence of which the best value is 
\[
\sigma_{DL} =2.49 \times 10^{-19} \text{ s} = 249 \text{zs}.
\]

\noindent Medium-term oscillations related to the instability of the pump make the OA deviation to increase for $t>80$ s. As expected from $t\sim600$ s the OA deviation for the single $\tau$ experiences the aforementioned long term drift, which is caused by temperature variations affecting both fibers and EOM response.

\noindent The differential detection ($\Delta \tau$), which is unaffected by pump instability or drifts, shows a saturation $S(\Delta \tau)>99.5 \%$ for averaging times $t \sim 10^2-10^3$ s.  Moreover, at $t=10^4$ s (blue circle), the total detection limit for differential measurements results 
\[
\sigma_{\Delta\tau}\approx 26 \text{ zs} .
\]
\section{Conclusions}
In this paper we implemented high resolution single photon Sagnac interferometry in a 2 km long single mode, polarization maintaining FOG with a quadrupole winding geometry. 

\noindent Our quantum-based delay sensor, operating in the telecom region and entirely fiber-coupled, shows the best sensitivity performances ever reported in literature for low photons regime Sagnac interferometry, capable to detect delays as low as $249$ zs at $72$ s integration time, and outperforming similar quantum sensors by more than an order of magnitude \cite{Lyons2018, sgobba2023attosecond}. The proposed experimental procedure proved to be the most resource-efficient way possible to measure a delay harnessing single photon quantum probability with a Sagnac interferometer since the differential measure $\Delta \tau$ is capable to achieve a Cramér-Rao bound saturation $S>99.5\%$ for integration times $t>10^2$s, and sensitivities as low as $26$ zs at averaging times $t=10^4$ s, corresponding to a detection limit on $\tau$ of $26/\sqrt{2} \approx 18$ zs in a set-up free from pump noise and drift. 
The detection limit of 26 zs  corresponds to a bias instability of a classical FOG operated with intense light of about  0.96  $^{\circ}/h$  proper of an intermediate grade FOG.
\noindent Since the employed FOG presents a total area of $125$ $\text{m}^2$, the aforementioned drift-free sensitivity at $t = 10^4$ s corresponds to a figure of merit $F=\frac{\sigma_\tau}{\mathcal{A}} \sim 1.4\times 10^{-16} \frac{\text{s}}{\text{km}^2}$. To fully appreciate the breadth of these results it is helpful to recall that Brady et al. \cite{Brady2021} estimated an upper bound on the detection limit per unit area $F\sim 10^{-15} \frac{\text{s}}{\text{km}^2}$ for the detection of Sagnac effect due to Earth rotation.

\noindent Hence, provided that the whole setup is mounted on an appositely designed  platform capable to tilt by a certain angle (similar to the one proposed in \cite{Restuccia2019, Fink2017}), it will possible to perform tests based on Sagnac effect due to earth rotation in single photon regime.

\subsection*{Funding}
 Agenzia Spaziale Italiana, Nonlinear Interferometry at Heisenberg Limit (NIHL) project (CUP  F89J21027890005);\\
This project has received funding from the European Defense Fund (EDF) under grant agreement EDF-2021-DIS-RDIS-ADEQUADE (n$^\circ$101103417). Funded by the European Union. Views and opinions expressed are however those of the author(s) only and do not necessarily reflect those of the European Union. Neither the European Union nor the granting authority can be held responsible for them; \\
PON Ricerca e Innovazione 2014-2020 FESR FSC (Project ARS01 00734 QUANCOM).\\
V. T. acknowledges support from the Air Force Office of Scientific Research (FA8655-23-1-7046).

\subsection*{Acknowledgments}
 
We wish to acknowledge Vincenzo Buompane and Graziano Spinelli for technical
support.

\subsection*{Data Availability}
Data underlying the results presented in this paper  may be obtained from the authors upon reasonable request.

\bibliographystyle{ieeetr}
\bibliography{sample}

\begin{thebibliography}{10}

\bibitem{yang2018review}
S.~Yang and G.~Zhang, ``A review of interferometry for geometric measurement,''
  {\em Measurement Science and Technology}, vol.~29, no.~10, p.~102001, 2018.

\bibitem{Abbott2016}
B.~Abbott and al., ``Observation of gravitational waves from a binary black
  hole merger,'' {\em Physical Review Letters}, vol.~116, p.~061102, Feb. 2016.

\bibitem{Hoghooghi2019}
N.~Hoghooghi, R.~J. Wright, A.~S. Makowiecki, W.~C. Swann, E.~M. Waxman,
  I.~Coddington, and G.~B. Rieker, ``Broadband coherent cavity-enhanced
  dual-comb spectroscopy,'' {\em Optica}, vol.~6, p.~28, Jan. 2019.

\bibitem{Coddington2009}
I.~Coddington, W.~C. Swann, L.~Nenadovic, and N.~R. Newbury, ``Rapid and
  precise absolute distance measurements at long range,'' {\em Nature
  Photonics}, vol.~3, p.~351–356, May 2009.

\bibitem{sagnac1913ether}
G.~Sagnac, ``L'{\'e}ther lumineux d{\'e}montr{\'e} par l'effet du vent relatif
  d'{\'e}ther dans un interf{\'e}rom{\`e}tre en rotation uniforme,'' {\em CR
  Acad. Sci.}, vol.~157, pp.~708--710, 1913.

\bibitem{Arianfard2023}
H.~Arianfard, S.~Juodkazis, D.~J. Moss, and J.~Wu, ``Sagnac interference in
  integrated photonics,'' {\em Applied Physics Reviews}, vol.~10, Feb. 2023.

\bibitem{Lefev2014}
H.~C. Lefèvre, ``The fiber-optic gyroscope, a century after sagnac’s
  experiment: The ultimate rotation-sensing technology?,'' {\em Comptes Rendus
  Physique}, vol.~15, p.~851–858, Dec. 2014.

\bibitem{Chow1985}
W.~W. Chow, J.~Gea-Banacloche, L.~M. Pedrotti, V.~E. Sanders, W.~Schleich, and
  M.~O. Scully, ``The ring laser gyro,'' {\em Reviews of Modern Physics},
  vol.~57, p.~61–104, Jan. 1985.

\bibitem{culshaw2005fiber}
B.~Culshaw, ``Fiber-optic sensors: applications and advances,'' {\em Optics and
  photonics news}, vol.~16, no.~11, pp.~24--29, 2005.

\bibitem{culshaw2005optical}
B.~Culshaw, ``The optical fibre sagnac interferometer: an overview of its
  principles and applications,'' {\em Measurement Science and Technology},
  vol.~17, no.~1, p.~R1, 2005.

\bibitem{lefevre2012fiber}
H.~Lefevre, ``The fiber-optic gyroscope: Achievement and perspective,'' {\em
  Gyroscopy and Navigation}, vol.~3, no.~4, pp.~223--226, 2012.

\bibitem{DellOlio2023}
F.~Dell’Olio, T.~Natale, Y.-C. Wang, and Y.-J. Hung, ``Miniaturization of
  interferometric optical gyroscopes: A review,'' {\em IEEE Sensors Journal},
  vol.~23, p.~29948–29968, Dec. 2023.

\bibitem{Hadfield2023}
R.~H. Hadfield, J.~Leach, F.~Fleming, D.~J. Paul, C.~H. Tan, J.~S. Ng, R.~K.
  Henderson, and G.~S. Buller, ``Single-photon detection for long-range imaging
  and sensing,'' {\em Optica}, vol.~10, p.~1124, Aug. 2023.

\bibitem{dello2022advances}
S.~Dello~Russo, A.~Elefante, D.~Dequal, D.~K. Pallotti, L.~Santamaria~Amato,
  F.~Sgobba, and M.~Siciliani~de Cumis, ``Advances in mid-infrared
  single-photon detection,'' in {\em Photonics}, vol.~9, p.~470, MDPI, 2022.

\bibitem{branning2000simultaneous}
D.~Branning, A.~L. Migdall, and A.~Sergienko, ``Simultaneous measurement of
  group and phase delay between two photons,'' {\em Physical Review A},
  vol.~62, no.~6, p.~063808, 2000.

\bibitem{dauler1999tests}
E.~Dauler, G.~Jaeger, A.~Muller, A.~Migdall, and A.~Sergienko, ``Tests of a
  two-photon technique for measuring polarization mode dispersion with
  subfemtosecond precision,'' {\em Journal of research of the National
  Institute of Standards and Technology}, vol.~104, no.~1, p.~1, 1999.

\bibitem{Lyons2018}
A.~Lyons, G.~C. Knee, E.~Bolduc, T.~Roger, J.~Leach, E.~M. Gauger, and
  D.~Faccio, ``Attosecond-resolution hong-ou-mandel interferometry,'' {\em
  Science Advances}, vol.~4, May 2018.

\bibitem{triggiani2023freq}
D.~Triggiani, G.~Psaroudis, and V.~Tamma, ``Ultimate quantum sensitivity in the
  estimation of the delay between two interfering photons through
  frequency-resolving sampling,'' {\em Phys. Rev. Appl.}, vol.~19, p.~044068,
  Apr 2023.

\bibitem{sgobba2023attosecond}
F.~Sgobba, A.~Andrisani, S.~Dello~Russo, M.~Siciliani~de Cumis, and
  L.~Santamaria~Amato, ``Attosecond-level delay sensing via temporal quantum
  erasing,'' {\em Sensors}, vol.~23, no.~18, p.~7758, 2023.

\bibitem{Park2019}
J.~Park, H.~Kim, and H.~S. Moon, ``Polarization-entangled photons from a warm
  atomic ensemble using a sagnac interferometer,'' {\em Physical Review
  Letters}, vol.~122, Apr. 2019.

\bibitem{Grace2020}
M.~R. Grace, C.~N. Gagatsos, Q.~Zhuang, and S.~Guha, ``Quantum-enhanced
  fiber-optic gyroscopes using quadrature squeezing and continuous-variable
  entanglement,'' {\em Physical Review Applied}, vol.~14, Sept. 2020.

\bibitem{Zych2012}
M.~Zych, F.~Costa, I.~Pikovski, T.~C. Ralph, and {\v{C}}.~Brukner, ``General
  relativistic effects in quantum interference of photons,'' {\em Classical and
  Quantum Gravity}, vol.~29, p.~224010, Oct. 2012.

\bibitem{Restuccia2019}
S.~Restuccia, M.~Toro{\v{s}}, G.~M. Gibson, H.~Ulbricht, D.~Faccio, and M.~J.
  Padgett, ``Photon bunching in a rotating reference frame,'' {\em Physical
  Review Letters}, vol.~123, Sept. 2019.

\bibitem{Bertocchi2006}
G.~Bertocchi, O.~Alibart, D.~B. Ostrowsky, S.~Tanzilli, and P.~Baldi,
  ``Single-photon sagnac interferometer,'' {\em Journal of Physics B: Atomic,
  Molecular and Optical Physics}, vol.~39, pp.~1011--1016, Feb. 2006.

\bibitem{Fink2019}
M.~Fink, F.~Steinlechner, J.~Handsteiner, J.~P. Dowling, T.~Scheidl, and
  R.~Ursin, ``Entanglement-enhanced optical gyroscope,'' {\em New Journal of
  Physics}, vol.~21, p.~053010, May 2019.

\bibitem{Zhou2003}
C.~Zhou, G.~Wu, L.~Ding, and H.~Zeng, ``Single-photon routing by time-division
  phase modulation in a sagnac interferometer,'' {\em Applied Physics Letters},
  vol.~83, pp.~15--17, July 2003.

\bibitem{Rideout2012}
D.~Rideout, T.~Jennewein, G.~Amelino-Camelia, T.~F. Demarie, B.~L. Higgins,
  A.~Kempf, A.~Kent, R.~Laflamme, X.~Ma, R.~B. Mann,
  E.~Mart{\'{\i}}n-Mart{\'{\i}}nez, N.~C. Menicucci, J.~Moffat, C.~Simon,
  R.~Sorkin, L.~Smolin, and D.~R. Terno, ``Fundamental quantum optics
  experiments conceivable with satellites{\textemdash}reaching relativistic
  distances and velocities,'' {\em Classical and Quantum Gravity}, vol.~29,
  p.~224011, Oct. 2012.

\bibitem{Hilweg2017}
C.~Hilweg, F.~Massa, D.~Martynov, N.~Mavalvala, P.~T. Chru{\'{s}}ciel, and
  P.~Walther, ``Gravitationally induced phase shift on a single photon,'' {\em
  New Journal of Physics}, vol.~19, p.~033028, Mar. 2017.

\bibitem{Brady2021}
A.~J. Brady and S.~Haldar, ``Frame dragging and the hong-ou-mandel dip:
  Gravitational effects in multiphoton interference,'' {\em Physical Review
  Research}, vol.~3, Apr. 2021.

\bibitem{Serrodyne}
H.~C. Lefevre, {\em The Fiber-optic Gyroscope}.
\newblock Third. Ed., Artech House, USA, Feb. 2014.

\bibitem{kay1993statistical}
S.~M. Kay, ``Statistical signal processing: estimation theory,'' {\em Prentice
  Hall}, vol.~1, pp.~Chapter--3, 1993.

\bibitem{Harnchaiwat2020}
N.~Harnchaiwat, F.~Zhu, N.~Westerberg, E.~Gauger, and J.~Leach, ``Tracking the
  polarisation state of light via hong-ou-mandel interferometry,'' {\em Optics
  Express}, vol.~28, p.~2210, Jan. 2020.

\bibitem{Cramer1999}
H.~Cram{\'e}r, {\em Mathematical methods of statistics}, vol.~9.
\newblock Princeton university press, 1999.

\bibitem{riley2008handbook}
H.~D.~A. Sullivan D~B, Allan D~W and W.~F. L, ``Characterization of clocks and
  oscillators natl. inst. stand. technol. technical note 1337,'' {\em
  Characterization of clocks and oscillators Natl. Inst. Stand. Technol.
  Technical Note 1337}, 1990.

\bibitem{Fink2017}
M.~Fink, A.~Rodriguez-Aramendia, J.~Handsteiner, A.~Ziarkash, F.~Steinlechner,
  T.~Scheidl, I.~Fuentes, J.~Pienaar, T.~C. Ralph, and R.~Ursin, ``Experimental
  test of photonic entanglement in accelerated reference frames,'' {\em Nature
  Communications}, vol.~8, May 2017.

\end{thebibliography}

\end{document}